\documentclass[figures,cite,nobm]{epl}
\usepackage{epsf}
\title{Influence of nano-mechanical properties on single electron tunneling:
  A vibrating Single-Electron Transistor}
\shorttitle{A vibrating Single-Electron Transistor}
\author{Daniel Boese\inst{1,2}\thanks{E-mail:
  \email{dboese@tfp.physik.uni-karlsruhe.de}} and Herbert Schoeller\inst{2,1}}
  \institute{ 
\inst{1}Institut f\"ur Theoretische Festk\"orperphysik, Universit\"at
  Karlsruhe, D-76128 Karlsruhe, Germany\\
\inst{2}Forschungszentrum Karlsruhe, Institut f\"ur Nanotechnologie, D-76021
Karlsruhe, Germany}
\date{\today}
\begin{document}
\maketitle
\begin{abstract}
We describe single electron tunneling through molecular structures under the
influence of nano-mechanical excitations. We develop a full quantum mechanical
model, which includes charging effects and dissipation, and apply it to the
vibrating C$_{60}$ single electron transistor experiment by Park {\em et al.}
{[Nature {\bf 407}, 57 (2000)].} We find good agreement and argue 
vibrations to be essential to molecular electronic systems. We propose a 
mechanism to realize negative differential conductance using local bosonic
excitations.
\end{abstract}
{\em Introduction.} 
Experiments on electronic transport through nano-scale systems show a
variety of physical conduction mechanisms. Due to their small
size quantum mechanics becomes important and Coulomb blockade, interference,
and Kondo physics are observed. Molecular systems
\cite{persson-etal,bezryadin-etal,klein-etal,reed-etal,zhou-etal,chen-etal,porath2-etal,kergueris-etal}
are characterized by large electronic energies beyond room temperature
and therefore offer the possibility to measure positions of molecular 
orbitals by transport spectroscopy. Furthermore, 
mechanical degrees of freedom in molecules have energies of order 
$1-10$ meV which can be probed experimentally for temperatures in the 
Kelvin regime. This letter addresses the latter topic and combines 
nano-electronic with nano-mechanical properties, in particular we model the
experiment in Ref.~\cite{park}. We find that local bosonic excitations
have an important influence on single electron tunneling,
and thus need to be included in models for molecular
electronics. Based on those systems we propose another way to realize negative
differential conductance (NDC). 

Attempts to model transport through molecular nano structures up to now
focused more on the electronic structure\cite{diventra}. In this letter we go
one step further and include molecular vibrations (or any other local bosonic
excitation) as well.
We remark that our approach is fundamentally different from models where the
vibration serves as a shuttle for the electrons\cite{gorelik,erbe}.
We address the case where the vibrational frequency is several orders of
magnitude larger than the frequency associated with tunneling events. 
Moreover the physics is dominated by charging effects, therefore the 
single charge tunneling picture\cite{curacao} is more appropriate than the 
scattering picture\cite{landauer}.
 
{\em Experiment.}
In their experiment Park and coworkers measured the current through a single
C$_{60}$ molecule. Using a break junction technique
they placed the molecule between two gold electrodes, to which it is weakly
bound, {\em i.e.}~the molecule sits in a Lennard-Jones-like potential on the
gold surface.  They could vary the source-drain and a gate voltage, that 
shifted the lowest unoccupied molecular orbital (LUMO). Thus they
have built a C$_{60}$ single electron transistor (SET). 
We briefly summarize their findings. The I-V curve displayed a large
conductance gap, which could be varied with the gate voltage. Once the current
was flowing a series of equally spaced ($\omega_0\approx 5$~meV) little jumps
followed, which was attributed to
the vibrational excitation of the whole molecule (center of mass motion) in the
Lennard-Jones potential discussed above. More excitations at higher energies
could also be found and could be attributed to {\em e.g.}~intra-molecular
vibrations.  
The experiment was conducted at $T\cong 1.5$~K $ = 0.13$~meV, the
charging energy exceeded $100$~meV, and the separation of the LUMOs is assumed
to be of similar magnitude. From the current one can deduce the tunneling
broadening $\Gamma$ to be of ${\mathcal{O}} (0.1$~$\mu$eV$)$.  

\begin{figure}
  \onefigure[scale=0.33]{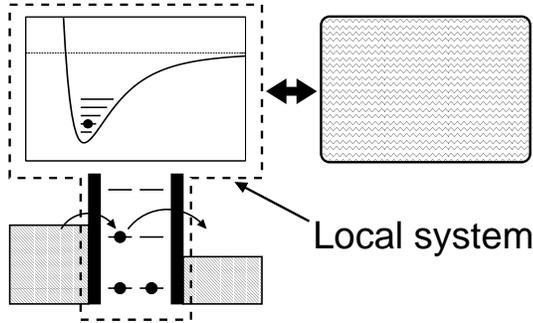}
        \caption{Theoretical model system: A local system consisting of a
          quantum dot and a harmonic oscillator is coupled to two electronic
          reservoirs and a boson bath.} 
        \label{fig:sys}
\end{figure}

{\em Theoretical Model.} 
Our model describes the dynamics of a local system, consisting of
the electronic states and a single-mode bosonic excitation due to the harmonic
potential, which accounts for the 
vibrations. The local system is coupled to two electronic reservoirs, a
boson bath, which allows for dissipation and a gate that can shift the
position of the MOs, see Fig.~\ref{fig:sys}. Hence the
Hamiltonian reads
\begin{equation}
H = H_{\mathrm{eres}} + H_{\mathrm{ph}} + H_{\mathrm{mol}} +
H_{\mathrm{mol-ph}} + H_{\mathrm{mol-eres}}.
\end{equation}
The electronic reservoirs are described by non-interacting electrons with a
constant density of states $\rho_0$, 
hence $H_{\mathrm{eres}} = \sum_{k \sigma r}
\epsilon_{k \sigma r} a_{k \sigma r}^\dagger a_{k \sigma r}$, 
where $r \in \{ \mathrm{L,R} \}$ labels the left and right electrode and
$\sigma$ the spin. The bosonic reservoir is given by 
$H_{\mathrm{ph}}=\sum_q\omega_q d_q^\dagger d_q$. The charging energy and the
separations of the MOs is large enough to take the LUMO as the
only participating MO. For the following discussion we take the LUMO to be
spin-degenerate, and include the Coulomb interaction in a 
Hubbard-like charging term. With
a projection operator for the vibrational states
$P_{ll'}=\left|l\right\rangle\left\langle l' \right|$ we have 
\begin{equation}
H_{\mathrm{mol}} = \sum_\sigma \epsilon_\sigma (V_g) c_\sigma^\dagger c_\sigma
+ U n_\uparrow n_\downarrow + \sum_l \epsilon_l^{\mathrm{vib}} P_{ll} .
\end{equation}
The gate electrode is hidden in the dependence of $\epsilon_\sigma$ on the
gate's potential. The energy levels of the oscillator are
$\epsilon_l^{\mathrm{vib}} = (l +1/2) \omega_0$. The coupling to the boson
bath ensures relaxation and dissipation of the vibrational energy in a 
natural way, and is generically written as
\begin{equation}
H_{\mathrm{mol-ph}} = \sum_{qll'} g_{ll'q} P_{ll'} \left(
  d_q + d_q^\dagger \right). 
\end{equation}
For simplicity, the constants $g_{ll'q}$ are assumed to be independent 
of $l,l'$ and we introduce the coupling $\alpha$ via the spectral density 
$\alpha |\omega |^{s} = 2\pi \sum_q |g_q|^2 \delta (\omega - \epsilon_q)$. 
The exponent $s$ depends on the underlying microscopic realization of the 
boson bath, i.e. acoustical phonons, fluctuations in the electro-magnetic 
environment, etc. In the spirit of the Caldeira-Legett model\cite{legett}, 
we assume from now on an Ohmic bath, {\em i.e.}~$s=1$. Finally the tunneling
term, which includes the possibility to excite the whole C$_{60}$ molecule 
to a vibration against the surface during a tunneling event, 
\begin{equation}
H_{\mathrm{mol-eres}} = \sum_{k\sigma r ll'} T^r \tau_{ll'} 
P_{ll'}\left( a_{k\sigma r}^\dagger c_\sigma + \mathrm{h.c} \right).
\end{equation}
Here, $T^r$ describes the tunnel matrix element in the absence of 
vibrational states, and the factor $\tau_{ll'}$ accounts for vibrational
excitations. Their microscopic origin can be understood\cite{park}
similar to what is known as the Franck-Condon principle: The electron tunnels
faster than the molecule can relax into a new state. When one electron tunnels
onto the molecule, it is likely that the charge distribution around it,
{\em i.e.}~the image charges in the gates which serve as capacitors, is not
uniform. Consequently there will be a force acting upon the charged
C$_{60}$. Hence the potential minimum $x_0$ is shifted slightly by a distance
$\delta$. Therefore we need the matrix elements between the states 
$\psi_{x_0,l'}$ in the potential centered at $x_0$ and the states 
$\psi_{x_0+\delta,l}$ of the potential shifted by the distance $\delta$, 
{\em i.e.}~$\tau_{ll'}\sim\langle\psi_{x_0+\delta,l} | \psi_{x_0,l'}\rangle$. 
For the system under consideration the states are well described by the 
harmonic oscillator ones. Obviously it gives $\delta_{l,l'}$ for 
$\delta=0$ or zero for $\delta$ large, and we find
that only a few transitions have considerable weight. With
the tunneling broadening $\Gamma^r = 2 \pi \rho_0 |T^r|^2$, we introduce 
the couplings $\Gamma_{ll'}^r= \Gamma^r |\tau_{ll'}|^2$ and 
$\Gamma_{ll'}=\sum_r\Gamma_{ll'}^r$, which control 
both the current and the excitations. 

{\em Theory.} 
The dynamics of the LUMO and all local bosonic excitations can be
captured by a kinetic equation for the reduced density matrix, 
which is obtained by tracing out
the reservoir and bath degrees of freedom of the full density matrix. 
For the parameters of interest double
occupancy and off-diagonal matrix elements can be neglected ($E_C \gg
\{T,\Gamma,\omega_0\}$ and $\omega_0 \gg \Gamma$).  Then only
the rates describing electron tunneling including vibrational
excitation, and those related to bosonic transitions are required. Note that
they are 
completely distinct, since the charge on the molecule changes by one for the
electronic ones, and remains constant for the boson-mediated ones. For
weak tunneling $\Gamma\ll T$ and weak dissipation $\alpha\ll 1$, we can perform
perturbation theory in the coupling constant in both cases. The rate
describing a transition from zero charge and vibrational state $l'$ to charge
one with spin $\sigma$ and vibrational state $l$ is given by 
$\Sigma_{\sigma l,0 l'}=\sum_r\Sigma^r_{\sigma l, 0 l'}$ with 
\begin{equation}
\Sigma_{\sigma l, 0 l'}^r = \Gamma_{ll'}^r f_r \left( \epsilon_\sigma 
+ \epsilon_l^{\mathrm{vib}} - \epsilon_{l'}^{\mathrm{vib}}\right).
\end{equation} 
For the reverse process the Fermi function $f_r (\omega)$ will be replaced by
$(1-f_r(\omega))$. The rate describing phonon absorption and an accompanying
vibrational excitation from $l'$ to $l$ (without change of the electronic
state) is given by 
\begin{equation}
\Sigma_{l,l'}^+ = \alpha \left| l'-l \right| \omega_0 N\left(
\epsilon_l^{\mathrm vib}-  \epsilon_{l'}^{\mathrm vib} \right).
\end{equation}
For the emission rate the Bose function $N(\omega)$ is replaced by
$1+N(\omega)$. The kinetic equation for the probabilities of the state 
$s$ reads
$\dot{P}_s =0 =\sum_{s'} (\Sigma_{ss'} P_{s'} - \Sigma_{s's}
P_{s})$, and the current is obtained via
\begin{equation}
I = e \sum_{\sigma,l,l'} \left( P_{0 l} \Sigma_{\sigma l', 0 l}^L -  P_{\sigma
    l'} \Sigma_{0 l, \sigma l'}^L  \right). 
\end{equation}

\begin{figure}
  \centerline{\includegraphics[width=8cm]{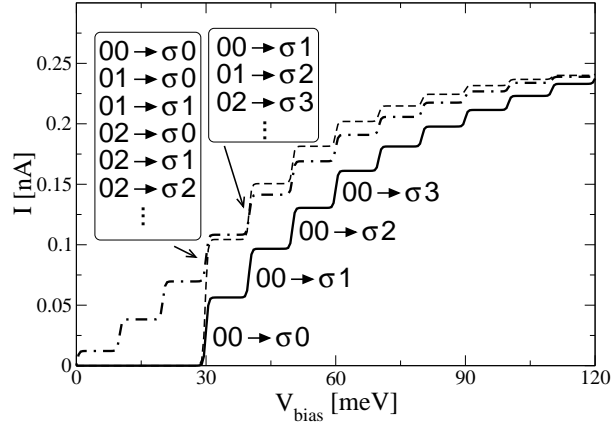}}
        \caption{Influence of the boson bath on the I-V characteristics:
          Emission only (solid, $T_{\mathrm{ph}}=0.13$ meV),
          no phonon bath (dashed), and emission and absorption 
          (dot-dashed, $T_{\mathrm{ph}}=13$ meV).
          $T_{\mathrm{el}}=0.13$ meV, $\epsilon_\sigma=15$ meV, 
          $\alpha=0.05$, $\omega_0=5$ meV, $\Gamma^r=0.015$ $\mu$eV, 
          $10$ vibrational states. Transitions are indicated for the 
          solid and dashed line. } 
        \label{fig:bath} 
\end{figure}

{\em Results and Discussion.}
Before we focus on the specifics we discuss the general
properties of the I-V characteristics. We tune the gate voltage such that the
LUMO is above the Fermi levels. The voltage is applied symmetrically,
{\em i.e.}~$\mu_{L,R}=\pm eV/2$. For low boson temperatures (solid curve
in Fig.~\ref{fig:bath}), absorption of phonons is suppressed and 
spontaneous emission, which in many molecules 
happens on a shorter time scale than the electron tunneling\cite{salam}, 
will force the molecule to stay predominantly in its vibrational ground state.
In this case, as in the experiment, we see no current until 
$e V_{\mathrm{bias}} = 2 \epsilon_\sigma$ followed by a series
of small steps with distance $2 \omega_0$. The latter are due to
excitations of higher vibrational states while tunneling. Each
time the bias voltage exceeds $2(\epsilon_\sigma+n\omega_0)$, a
new tunneling-in process $|0,0\rangle \rightarrow |\sigma,n\rangle$ 
can occur leading to an increase of the current. The excited vibrational 
state relaxes immediately to the ground state $n=0$ and tunneling-off 
happens via the process $|\sigma,0\rangle \rightarrow |0,0\rangle$. 
The number of steps is determined by the number of vibrational states 
taken into account. The width of the rise for each single step is 
controlled by $\mathrm{max}(T_{\mathrm{el}}, \Gamma)$, which in our case is 
$T_{\mathrm{el}}$. In Fig.~\ref{fig:bath} we show results where all
vibrational states can be accessed equally, {\em i.e.}~$\tau_{ll'}=1$. 
We remark that our results for strong dissipation are 
not special for the Ohmic bath, but rather independent of (positive) 
$s$, since the energy is dissipated immediately.

The situation is more complicated in the absence of dissipation 
(dashed curve in Fig.~\ref{fig:bath}). Here, once the threshold for the
onset of current has been reached, the tunneling-off processes 
$|\sigma,0\rangle \rightarrow |0,n\rangle$, with
$n\omega_0 < \epsilon_\sigma$, can leave the 
initially not excited molecule in an excitated state. This state
then serves as a starting point for other tunneling-in processes
$|0,n\rangle \rightarrow |\sigma,m\rangle$, with $m\le n$, leading 
to an increase of the current compared to the case with dissipation
(after each additional step in the I-V-characteristics, the allowed
values for $m$ increase by one).

For high boson temperatures where emission and absorption of phonons
is possible (dot-dashed curve in Fig.~\ref{fig:bath}), the I-V curve
for $e V_{\mathrm{bias}} > 2 \epsilon_\sigma$ is similiar to the
case without dissipation. However, below the treshold, current
precursors occur due to tunneling-in processes where the 
vibrational state lowers its energy. The higher vibrational states
have here a finite probability due to absorption of phonons which
is not possible in the absence of dissipation. However, since this 
is not seen in the experiment, one can rule out such high boson 
temperatures. 
 
Experimentally the number of steps that can be resolved is much less than $10$
and we therefore proceed by taking into account the dependence of the
Franck-Condon factor $\tau_{ll'}$ on the vibrational states. Assuming that 
the potential is well described in a harmonic approximation with parameters 
from the Lennard-Jones potential\cite{lennard}, then
only the shift $\delta$ is unknown. A crude estimate, based on the force due
to an image charge in the electrode, is given by Park {\em et al} \cite{park}.
Going beyond this picture is difficult because the electrodes' geometry
is not known. However, our results show that a better estimate may be
obtained. In Fig.~\ref{fig:fc} we show the I-V characteristics for three 
$\delta$'s of the order of the previous estimate of $4$ pm. 
The ratio of the first and second step height is highly
sensitive to $\delta$. Comparing with the experiment one therefore would
suggest a value of $\delta \approx 3$ pm. Moreover it can be seen that the
current saturates much faster than without Franck-Condon factors. This is not
surprising, because the molecule is mostly in its ground state from which the
higher states can not be reached. This is different without dissipation, where
even a small NDC can be realized.
The differential conductance, which shows a peak for every excitation, is
shown in Fig.~\ref{fig:dIdV}, with the NDC effect in the
inset. 
\begin{figure}
 \twofigures[width=7cm]{fc_delta.eps}{dIdV.eps}
        \caption{Including the Franck-Condon factors with $\delta=3$ pm
          (solid), $\delta=4$ pm (dashed), and  $\delta=5$ pm (dot-dashed). 
          Inset: Comparison of with (lower) and without Franck-Condon factors 
          (upper curve). $T_{\mathrm{el}}=T_{\mathrm{ph}}=0.13$meV, 
          $\epsilon_\sigma=15$ meV,  $\alpha=0.05$, $\omega_0=5$ meV,
          $\Gamma^r=0.1$ $\mu$eV, $10$ vibrational states.}  
        \label{fig:fc} 
        \caption{dI-dV curve, $\delta=3$ pm with bath (solid) and without
          (dashed in inset), and $\delta=4$ pm with bath
          (dot-dashed). $T_{\mathrm{el}}=T_{\mathrm{ph}}=0.13$ meV
          $\epsilon_\sigma=15$ meV,  $\alpha=0.05$, $\omega_0=5$ meV,
          $\Gamma^{0,r}=0.1$ $\mu$eV, $10$ vibrational states } 
        \label{fig:dIdV}
\end{figure}
We remark that our results  
agree very well with the experimental data\cite{park}. However, there remain
features which can not be reproduced in our calculation, namely a rising
background and more excitations at higher energies. The origin of the
background is not precisely known yet, the other excitations will be addressed
below.

The NDC effect as seen before is due to a new mechanism specific for 
additional bosonic degrees of freedom. To exhibit the physics, 
Fig.~\ref{fig:ndcm} shows the effect for a simplified model of one
electronic level with two vibrational states $l=0,1$.
The $T=0$ currents of the two plateaus are (for $\Gamma_{01}=\Gamma_{10}$)
\begin{eqnarray}
I_{\mathrm{mid}} &=& P_{00} \Gamma_{00} + P_{01}(\Gamma_{01}+\Gamma_{11}) -
P_{11} \Gamma_{01} = 
\Gamma_{00} \frac{2 \Gamma_{01} (\Gamma_{00} +
  \Gamma_{01} + 2 \Gamma_{11}) + \Gamma_{11} (3 \Gamma_{00} + \Gamma_{11})} {6
  \Gamma_{01} \Gamma_{00} + 2\Gamma_{11} (\Gamma_{01} + 4 \Gamma_{00} )}
  \nonumber \\
I_{\mathrm{fin}} &=& P_{00} (\Gamma_{00}+ \Gamma_{01}) +
P_{01}(\Gamma_{01}+\Gamma_{11}) = 
\frac{1}{4} \left( \Gamma_{00} + 2 \Gamma_{01} +
  \Gamma_{11} \right).
\end{eqnarray}
In the limit $\Gamma_{01} \ll \Gamma_{00}$ we see that only 
$\Gamma_{11}<\Gamma_{00}$ is needed to observe NDC. For 
$\Gamma_{00}\gg\Gamma_{10}=\Gamma_{01}=\Gamma_{11}$, we get
$I_{\mathrm{fin}}/I_{\mathrm{mid}}=14/20$. The origin is that for the
second plateau the probability distribution changes such that even 
weakly coupled vibrational states are occupied equivalently to the 
strongly coupled ones. However, their contribution to the current is 
small, and therefore the current decreases. We stress that the 
$\tau_{ll'}$ naturally show the appropriate kind of behavior. Assuming 
a simple functional form for them we show in Fig.~\ref{fig:ndcm} how 
many bosonic levels can in principle lead to large peak to valley ratios. 
For multi-mode bosonic excitations the effect survives and can even be more 
drastic, when many
excitations are energy degenerate. NDC has already been seen in other
nanostructures\cite{johnson,weis,chen}, and explanations in terms of
spin-blockade\cite{weinmann} and asymmetric coupling\cite{hettler} have been
suggested. In our case however, no additional electronic states at higher
energies are required, nor do we need the full many-body state of the
structure. 
\begin{figure}
 \twofigures[scale=0.28]{ndc.eps}{ndc_many.eps}
        \caption{Negative differential conductance with two vibrational
          states. Main panel shows $\Gamma_{11}$ ranging from $0.01\,
          \Gamma_{00}$ (bottom curve) via $0.1\, \Gamma_{00}$,
          $0.2 \, \Gamma_{00}$, $0.5 \, \Gamma_{00}$, $0.7 \, \Gamma_{00}$,
          $0.8 \, \Gamma_{00}$  to $0.9 \Gamma_{00}$ (top
          curve). $\alpha=0$. Inset: $\alpha=5 \, 10^{-5}$ (top) down $5
          \, 10^{-9}$. Each line corresponds to one order of
          magnitude. Common parameters are $T_{\mathrm{el}}=T_{\mathrm{ph}}=0.13$
          meV, $\epsilon_\sigma=15$ meV, $\omega_0=5$ meV,
          $\Gamma_{01}=\Gamma_{10}=0.1\, \Gamma_{00}$ and
          $\Gamma^{r}=0.1$ $\mu$eV.} 
        \label{fig:ndc}
\caption{Negative differential conductance with $N$ vibrational
          states and $\tau_{ll'}=1/(|l-l'|+1)^2 1/((l+l')/2)^2$.
          Parameters are $T_{\mathrm{el}}=T_{\mathrm{ph}}=0.13$
          meV, $\epsilon_\sigma=2$ meV, $\omega_0=5$ meV, and
          $\Gamma^{r}=0.1$ $\mu$eV.}
\label{fig:ndcm}
\end{figure} 

Finally we comment on the influence of bosonic excitations in general 
molecular electronic devices. Multi-mode bosonic excitations can be 
included in the same way. Other degrees of freedom could be due to 
internal vibrations, {\em e.g.} the torsional
vibration of the oligo-phenyl molecules used in the experiment by Chen and
coworkers\cite{chen} have energies in the meV range and should be observable 
at low enough temperatures. The structure of smaller bosonic excitations would 
display itself on top of the larger ones, provided they
are well separated. For heavier molecules where $\omega_0\sim\Gamma$ 
we expect for $T<\Gamma$ an interesting interplay including 
interference effects between the
various excitations, partly related to off-diagonal elements of the density
matrix. In addition, for $T\gg\Gamma$, this can lead to an increasing 
background vibrational noise. This background (not in the gap) is indeed 
seen in the
experiments and it is quite likely that such small local bosonic excitations
play an essential role in it. We remark that this kind of behavior is
fundamentally different from artificial semiconductor nano structures, which
are embedded in a substrate.

{\em Summary.} We described single electron tunneling through a system with
local bosonic degrees of freedom and dissipation to external heat baths.
We applied our model to a C$_{60}$ molecule, which can be excited to center of
mass 
oscillations. We found that our results compare well with the 
experiment. We showed that this model may be used to realize negative 
differential conductance. We concluded that transport
through molecular nano-devices is inevitably controlled by its electronic
and mechanical structure.
\acknowledgments
We would like to thank Filipp Furche, Matthias Hettler, Silvia Kleff, Gerd
Sch\"on and Florian Weigend for valuable discussions. This work is supported
by the DFG as part of the Graduiertenkolleg ''Kollektive Ph\"anomene im
Festk\"orper'' (D.B.) and via "SFB 195" (D.B. and H.S.).


\begin{thebibliography}{99}
\bibitem{lennard}
The molecule sits in the combined potential of the two electrodes, which are
so close together that a binding energy of about $2$ eV can be assumed. The
minimum is about $6.2$ {\AA} away from the surface from which the energy
$\omega_0 \cong 5$ meV can be deduced. See also 
\Name{Chavy C., Joachim C. \and Altibelli A.}
\REVIEW{Chem. Phys. Lett.}{412}{1993}{569}; 
\Name{Ruoff R. S. \and Hickman A. P.}
\REVIEW{J. Phys. Chem.}{97}{1993}{2494}
\end{thebibliography}
\end{document}